\documentstyle[twoside,fleqn,espcrc2,epsf]{article}

\newcommand{\beq}{\begin{equation}}
\newcommand{\eeq}{\end{equation}}

\newcommand{\AmS}{{\protect\the\textfont2
  A\kern-.1667em\lower.5ex\hbox{M}\kern-.125emS}}

\title{Phase transition and topology in 4d simplicial gravity}

\author{S.~ Bilke\address{Fakult\"{a}t f\"{u}r Physik, Universit\"{a}t
Bielefeld, Postfach 10 01 31, Bielefeld 33501, Germany},
Z.~Burda$^{\mbox{\scriptsize a}}$\thanks{Permanent address: 
Institute of Physics, Jagellonian University, 
ul. Reymonta 4, 30-059 Krak\'{o}w, Poland}, 
A. Krzywicki\address{LPTHE, 
B\^{a}t. 211, Universit\'e de Paris-Sud, 91405 Orsay, France}
\thanks{Laboratoire associ\'e au CNRS.}
and B.~Petersson$^{\mbox{\scriptsize a}}$}

\begin{document}

\begin{abstract}
We present data indicating that the recent evidence 
for the phase transition being of first order does not 
result from a breakdown of the ergodicity of the algorithm. 
We also present data showing that the thermodynamical limit 
of the model is independent of topology.

\noindent
LPTHE Orsay 96/67, BI-TP 96/31, hep-lat/9608027
\end{abstract}

\maketitle

\section{Introduction}

When Euclidean Quantum Gravity is dis\-cretized following
the dynamical triangulation recipe, the Einstein-Hilbert 
action takes the form
\beq
S=\kappa_4 N_4 - \kappa_2 N_2
\eeq
where $N_i$ is the number of $i$-simplices and
$(\kappa_2,\kappa_4)$ are coupling constants. 
For a given topology the grand canonical 
partition function $Z_{GC}$ is defined by
\beq
Z_{GC}(\kappa_2, \kappa_4) = \sum_{\{ T\} } C(T) e^{-S}
\eeq
where $C(T)$ is a symmetry factor and the sum goes over 
simplicial complexes $T$. The canonical partition function 
$Z_C$ is defined by an analogous expression, 
where $N_4$ of $T$ is kept fixed.
One can investigate the critical behaviour of the 
theory by studying
\beq
F(\kappa_2,N_4)= \log Z_C(\kappa_2, N_4)
\eeq
and its derivatives in the limit $N_4 \rightarrow \infty$.
One finds that the model has two phases. The determination 
of the exact nature of the transition between 
these two phases is obviously of utmost importance.
\par
In Ref.~\cite{fo} it was found, in spherical topology, 
that the transition is of first order, contrary
to the common belief. 
However, one can wonder whether this finding 
does not reflect a deficiency of the simulation set-up. 
As it is an important question, we have decided to check 
it in more detail. The result of this investigation 
is presented in sect. 2 of this communication. 
In sect. 3 we also present some results on the behaviour of 
the model in other topologies.
 
\section{The dependence of the transition on the limitations
of the algorithm}

In 4d there does not exist a set of ergodic local moves
preserving the value of $N_4$. 
In order to study the canonical ensemble the standard 
procedure is to carry out  
a grand-canonical simulation with a modified action, 
for example
\beq
S_{mod}=  - \kappa_2 N_2  + \kappa_4^0 N_4 
+ \frac{\gamma}{2} (N_4 - V_4)^2
\label{kw}
\eeq
where $V_4$ is the value of $N_4$ to be studied. 
All five local moves ergodic in the grand-canonical ensemble
are employed. The constant $\kappa_4^0$ is chosen so that 
$N_4$ fluctuates around $V_4$: $\Delta N_4=1/\sqrt{\gamma}$. 
Measurements are performed every $n$'th time $N_4$ crosses $V_4$, 
with an appropriate choice of $n$. In Ref.~\cite{fo}  
we have used $\gamma=0.005$ and $0.0005$. 
The first-order nature of the transition was concluded 
from finite size scaling and from the appearance of 
a double-peak structure at $\kappa_2=1.258$ for our 
largest value of $N_4=32000$. With $\gamma=0.005$ ($0.0005$) 
one has $\Delta N_4= 14$ ($45$), a rather modest 
fluctuation compared to $32000$. 
However, the constraint limiting volume fluctuations 
may not be innocuous. Any two lattice configurations 
at $N_4=V_4$ can be connected by a sequence of local 
moves, provided $N_4$ is allowed to deviate enough 
from $V_4$. When this condition is satisfied, 
the path of the system, in its time evolution and in 
the space of configurations, recurrently crosses 
the hyperplane $N_4=V_4$. This defines the density of states, 
corresponding to the canonical ensemble. It is not obvious 
that this density is unaffected when volume 
fluctuations are constrained.
Therefore, we have performed another set of simulations 
at  $\kappa_2=1.258$ and $N_4=32000$, allowing for
fluctuations $\Delta N_4 \sim 10^4 $. One cannot
just set $\gamma=10^{-8}$: the finite size
corrections to $F(\kappa_2, N_4)$ lead to
metastable states at different values of $N_4$. Thus, 
instead of 
(\ref{kw}) we choose another action
\beq
S_{mod} = -\kappa_2 N_2 + U(N_4) \; ,
\eeq 
where the external potential $U$ is chosen so as to 
favour large excursions from the reference volume $V_4$.
The optimal choice is $U = -F(\kappa_2,N_4)$, 
but the free energy is unknown. Hence, we measured 
the derivative $\partial F/\partial N_4$ 
for a few volumes, interpolated it linearly and integrated 
over $N_4$. The result can
differ within the errors bars from the true free energy and 
this can make the simulation unstable. To 
insure stability we added a
constant attractive force $\delta = 0.0001$ to the 
derivative and reflecting walls at $16000$ and $64000$.
 \begin{figure}[t]
 \epsfxsize=75mm
 \epsfysize=58mm
   \epsfbox{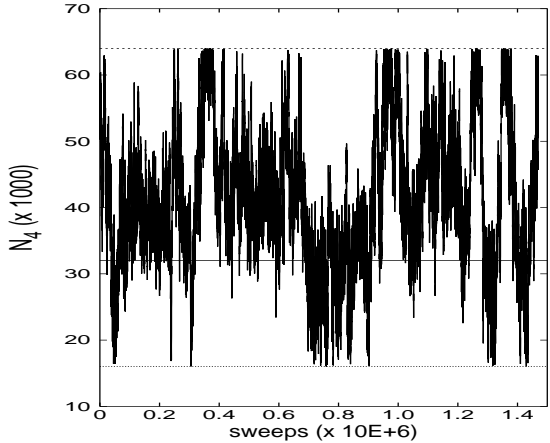}
  \vspace*{-5.0ex}
 \caption{The time history of $N_4$.}
 \end{figure}
The long excursions are evident in the history of $N_4$ as
shown in Fig.1. We have performed $1.5 \times 10^6$ 
sweeps and gathered $3 \times 10^5$ measurements, which 
were taken every 100'th crossing of $V_4$. 
In Fig.2 we show the corresponding $N_0$-distribution 
together with that of Ref.~\cite{fo}. They are in fact
very similar and, in particular, both are bimodal. 
The first three cumulants of $N_0$ distribution 
are now $0.1966(4)$, $0.318(26)$ and $-3.0(30)$ to be 
compared respectively with the values $0.1971(2)$, 
$0.316(8)$ and $-6.1(21)$ given in Ref.~\cite{fo}. They agree 
within the statistical errors.
 \begin{figure}[t]
\epsfxsize=75mm
\epsfysize=58mm
   \epsfbox{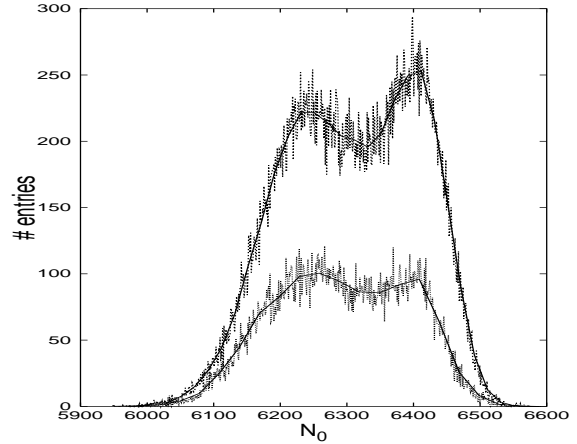}
  \vspace*{-5.0ex}
  \caption{The histograms of $N_0$.}
 \end{figure}
 \begin{figure}[t]
\epsfxsize=75mm
\epsfysize=83mm
   \epsfbox{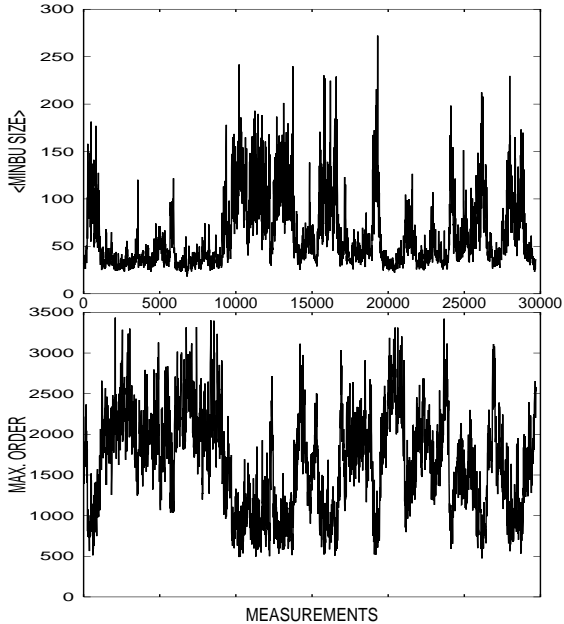}
  \vspace*{-5.0ex}
  \caption{The time history of the 
maximal vertex order and of the average minbu size.}
 \end{figure}
In Fig.3 we show the history of the average 
size of minimal neck baby universes, large in the 
elongated phase and small in the crumpled one, and the
order of the most singular point, 
which has the correlated opposite behaviour. 
The existence of two states with tunneling between them is very apparent. 
We conclude that the two state signal, 
i.e. the first-order nature of the transition is 
a genuine feature of the model.

\section{Dependence on topology}

We have investigated the dependence of the free energy on
topology. More precisely, we have measured
\beq                  
r=\frac{1}{N_4}\frac{\partial F}{\partial \kappa_2}=\frac{<N_2>}{N_4}   
\eeq
and
\beq
k(\kappa_2,N_4)=\frac{\partial F}{\partial N_4}
\eeq
In the limit $N_4 \rightarrow \infty$ they become intensive
quantities and $k(\kappa_2,N_4) \rightarrow \kappa_{4cr}$.
We have simulated the topologies $S^4$,$S^3 \times S^1$ and $(S^1)^4$.
The minimum configurations have $(N_{2MIN},N_{4MIN})$ equal
$(20,6)$,$(110,44)$ and $(1472,704)$ respectively. In the elongated phase
the most probable move, {\em ie} barycentric subdivision, leads to
\beq
\begin{array}{rl}
N_2 & =\frac{5}{2} N_4 + c \\ & \\
c & =N_{2MIN}-\frac{5}{2}N_{4MIN}
\end{array}
\eeq
and $c$'s are $5$,$0$ and $288$ for 
$S^4$,$S^3 \times S^1$ and $(S^1)^4$ respectively.
One expects
\beq
\begin{array}{rl}
r & =r_{\infty} + c N_4^{-1} + \dots  \\ & \\
k & =\kappa_{4cr} + (\gamma-3)N_4^{-1} + \dots 
\end{array}
\eeq
with $\gamma =\frac{1}{2}$. In Fig.4 we show $r$ as a
function of $1/N_4$ for $\kappa_2=2.0$. In fact, 
as one can see from figure 4 one gets a very good fit
by using $c$ from the minimum configurations.
%
 \begin{figure}[t]
\epsfxsize=75mm
\epsfysize=58mm
   \epsfbox{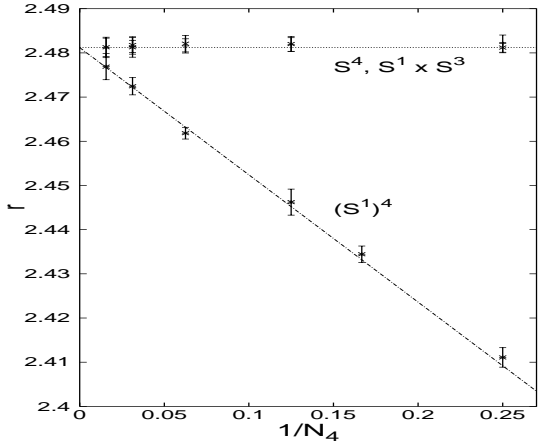}
  \vspace*{-5.0ex}
  \caption{The average action density $r$ vs. system size
  for various topologies.}
 \end{figure}
\par   
In the crumpled phase one expects
\beq
k = \kappa_{4cr} + f_k N_4^{-\delta} + \dots 
\eeq
\beq
r = r_{\infty}   + f_r N_4^{-\delta} + \dots
\eeq
where the coeeficients $f_k,f_r$ are functions of
$\kappa_2$ only. At $\kappa_2=0.0$ i.e. in the 
crumpled phase we have made
very extensive runs. We see no dependence on topology there, and
the best fit gives $\delta=0.49(12)$. This is in contrast to some earlier
results, and consistent with $\delta=1/2$. The latter value can be
obtained again from a purely kinematical argument, using the
fact that the most probable configurations have $N_4 \sim N_0^2$.
As a conclusion, the termodynamic limit is independent of topology,
and the finite size corrections can be understood from purely
kinamatic arguments. An extended version of our results on
topology dependence will be published in Ref.~\cite{top}.

We thank the CNRS computing center IDRIS and
the HLRZ J\"{u}lich for computer time.
Z.B. thanks the Deutsche Forschungsgemeinschaft 
for the financial support under Grant Pe 340/3-3.


\begin{thebibliography}{99}
\bibitem{fo} P. Bialas, Z. Burda, A. Krzywicki, B. Petersson,
Nucl. Phys. {\bf B472} (1996) 293.
\bibitem{top} S. Bilke, Z. Burda, B. Petersson, in preparation.
\end{thebibliography}
\end{document}